\documentclass{article}
\usepackage{spconf,amsmath,graphicx}
\usepackage{amssymb}
\usepackage{amsmath}
\usepackage{graphicx}
\usepackage{tcolorbox}
\usepackage{mathtools}
\usepackage{braket}

\usepackage[ruled]{algorithm2e}
\usepackage{bm}
\usepackage{romannum}
\usepackage{tabularx,ragged2e,booktabs,caption}
\usepackage{xcolor}
\usepackage{enumitem}
\usepackage{amsmath}
\usepackage[bookmarks=false]{hyperref}
\hypersetup{
    colorlinks=true,
    linkcolor=black,
    filecolor=black,      
    urlcolor=black,
}
\makeatletter
\newcommand{\norm}[1]{\left\lVert#1\right\rVert}
\newcommand{\E}{\operatorname*{\mathbb{E}}\ilimits@}
\makeatother

\DeclareMathOperator*{\argmin}{arg\,min}

\title{Predicting Flat-Fading Channels via Meta-Learned \\Closed-Form Linear Filters and 
Equilibrium Propagation}
\name{Sangwoo Park\qquad Osvaldo Simeone\thanks{The work of O. Simeone was supported by the European Research Council (ERC) under the European Union's Horizon 2020 research and innovation programme (grant agreement No. 725731).}}
\address{King's Centre for Learning \& Information Processing (KCLIP), \\ Department of Engineering, King's College London}
\begin{document}
%
\maketitle
\begin{abstract}
Predicting fading channels is a classical problem with a vast array of applications, including as an enabler of artificial intelligence (AI)-based proactive resource allocation for cellular networks. 
Under the assumption that the fading channel follows a stationary complex Gaussian process, as for Rayleigh and Rician fading models, the optimal predictor is linear, and it can be directly computed from the Doppler spectrum via standard linear minimum mean squared error (LMMSE) estimation. However, in practice, the Doppler spectrum is unknown, and the predictor has only access to a limited time series of estimated channels. This paper proposes to leverage meta-learning in order to mitigate the requirements in terms of training data for channel fading prediction. Specifically, it first develops an offline low-complexity solution based on linear filtering via a meta-trained quadratic regularization. 
Then, an online method is proposed based on gradient descent and equilibrium propagation (EP). Numerical results demonstrate the advantages of the proposed approach, showing its capacity to approach the genie-aided LMMSE solution with a small number of training data points.
\end{abstract}

\vspace{-0.4cm}
\section{Introduction}
\label{sec:intro}
\vspace{-0.3cm}
The capacity to accurately predict channel state information (CSI) is an enabler of efficient resource allocation strategies. Notably, it allows the transmitter of a communication link to optimize modulation and coding schemes, while reducing the pilot overhead; and it can be leveraged by artificial-intelligence (AI) tools to implement proactive networking protocols \cite{pqos_paper}. Given that standard fading channel models follow stationary complex Gaussian processes, e.g., for Rayleigh and Rician fading, optimal channel predictors can be obtained via linear minimum mean squared error (LMMSE) estimators, such as the Wiener filter \cite{hoeher1997two}. This, however, requires the channel statistics, including the Doppler spectrum, to be known. In practice, CSI statistics are a priori unknown, and need to be estimated from data.

Accordingly, there is a long line of work that focuses on predicting time-varying channels with linear predictors obtained via autoregressive (AR) models \cite{duel2000long} or Kalman filtering \cite{min2007mimo} by estimating channel statistics from available pilot data. Recently, deep learning-based nonlinear predictors have been proposed to adapt to channel statistics through the training of neural networks \cite{kim2020massive, jiang2020long}. However, as reported in \cite{kim2020massive, jiang2020long}, deep learning based predictors tend to require excessive training (pilot) data, while failing to outperform well-designed linear filters in the low-data regime.

In this paper, we propose to leverage meta-learning in order to further reduce the data requirement of linear CSI predictors for flat-fading channels. The key idea is to leverage multiple related channel data sets from various propagation environments to optimize the hyperparameters of a training procedure. Specifically, we first develop a low-complexity offline solution that obtains hyperparameters and predictors in closed form. Then, an online algorithm is proposed based on gradient descent and equilibrium propagation (EP) \cite{scellier2017equilibrium, zucchet2021contrastive}. Previous applications of meta-learning to communication systems include demodulation \cite{park2020learning, cohen2021learning},  channel equalization \cite{raviv2021meta}, encoding/decoding \cite{park2020meta,jiang2019mind, park2020end}, MIMO detection \cite{goutay2020deep}, beamforming \cite{yuan2020transfer, zhang2021embedding}, and resource allocation \cite{nikoloska2021black}.

\vspace{-0.4cm}
\section{System Model and Background}
\label{sec:model}
\vspace{-0.35cm}
\textit{Notation:} In this paper, $(\cdot)^\top $ denotes transposition; $(\cdot)^\dagger$ Hermitian transposition; $|\cdot|$ the absolute value; $||\cdot||$ the Euclidean norm; and $I$ the identity matrix.

\vspace{-0.53cm}
\subsection{Preliminaries}
\vspace{-0.23cm}
In order to establish some useful notation, consider a supervised learning problem targeting the prediction of a scalar label $y$ given a $d\times 1$ covariate vector $x$. Both $x$ and $y$ are complex-valued. The training data set $\mathcal{\mathcal{Z}} = (X,y)$ includes $n$ labelled examples $\{ (x_i, y_i) \}_{i=1}^n$. Specifically, the $n \times d$ matrix $X=[x_1^\dagger,\ldots,x_n^\dagger]^\top$ contains by row the Hermitian transpose of the $d \times 1$ covariate vectors $\{ x_i \}_{i=1}^n$, while the $n \times 1$ vector $y=[y_1,\ldots,y_n]^\dagger$ collects the complex conjugate of the corresponding labels $\{ y_i \}_{i=1}^n$. A linear predictor $\hat{y}=v^\dagger x$ of the label $y$ is defined by a $d\times 1$ model parameter vector $v$.

Given data set $\mathcal{\mathcal{Z}}$ and hyperparameters $(\lambda, \bar{v})$, with $\lambda > 0$ and $d \times 1$ bias vector $\bar{v}$, we define the solution of ridge regression as
\vspace{-0.5cm}
\begin{align}
\nonumber
v_\lambda^*(\mathcal{\mathcal{Z}} | \bar{v}) &= \argmin_{v} \left\{\sum_{i=1}^n | v^\dagger x_i - y_i|^2 + \lambda || v - \bar{v} ||^2 \right\} \\
&= \argmin_{v} \Big\{ || X v - y ||^2 + \lambda || v - \bar{v} ||^2 \Big\},
\label{eq:general_ridge}
\\[-0.9cm]\nonumber
\end{align}
which can be obtained explicitly as 
\vspace{-0.2cm}
\begin{align}
v_\lambda^*(\mathcal{\mathcal{Z}} | \bar{v}) = A^{-1} \big( X^\dagger y + \lambda \bar{v} \big), \textrm{ with } A = X^\dagger X + \lambda I.
\label{eq:general_ridge_sol}
\end{align}
\vspace{-1.13cm}
\subsection{System Model and Problem Definition}
\vspace{-0.2cm}
\label{sec:conven_channel_pred}
\begin{figure}
    \centering
    \hspace{-0cm}
    \includegraphics[width=\columnwidth]{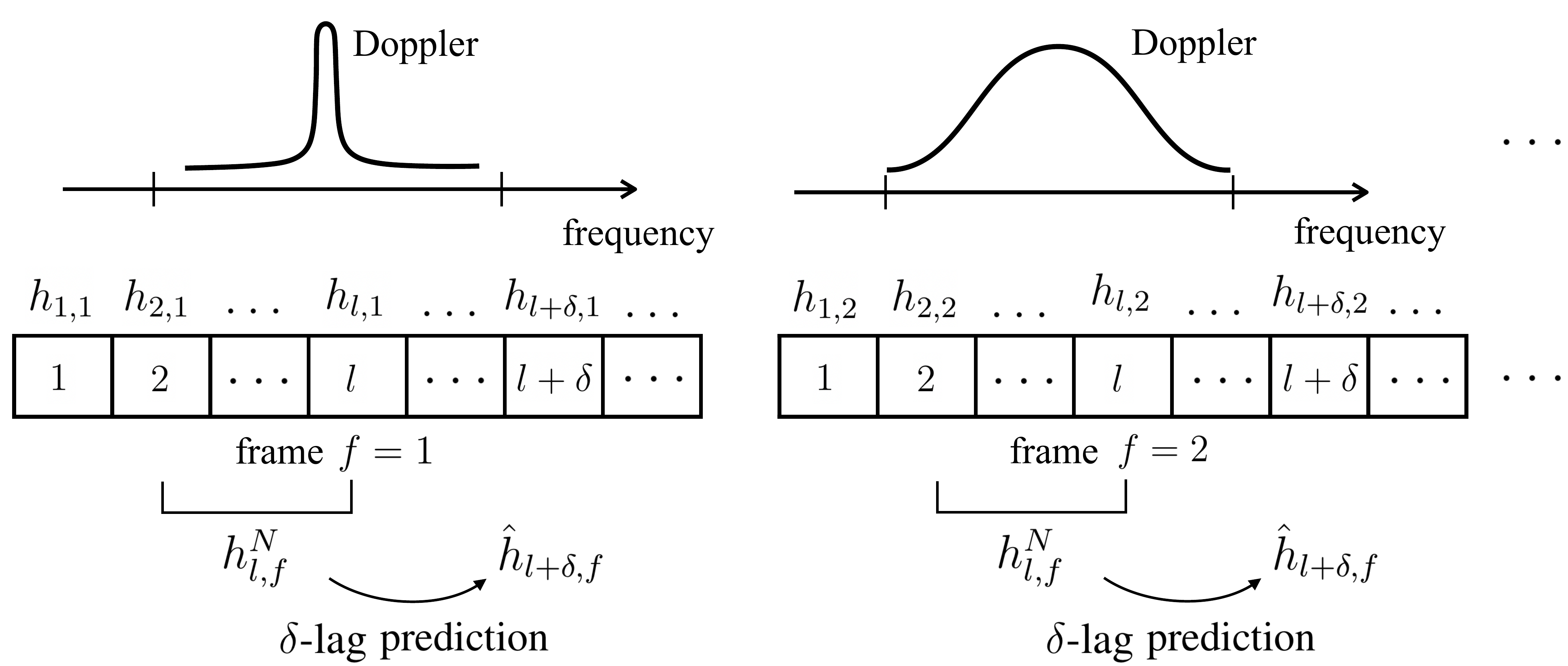}
    \caption{Illustration of the frame-based transmission system under study: At any frame $f$, based on the previous $N$ channels $h_{l,f}^N$, we investigate the problem of optimizing the $\delta$-lag prediction $\hat{h}_{l+\delta,f}.$}
    \label{fig:system_model}
    \vspace{-0.5cm}
\end{figure}

As illustrated in Fig.~\ref{fig:system_model}, we study a frame-based transmission system, with each frame containing multiple time slots. In each frame $f=1,2,\ldots$, the channels $h_{l,f} \in \mathcal{C}$ follow a stationary process across the time slots $l=1,2,\ldots,$ with a generally distinct Doppler spectrum for each frame $f$. For instance, in a frame $f$, we may have a slow-moving user subject to time-invariant fading, while in another the channel may have fast temporal variations described by Clarke's model with a large Doppler frequency. 

For each frame $f$, we are interested in addressing the lag-$\delta$ channel prediction problem, in which channel $h_{l+\delta,f}$ is predicted based on the $N$ past channels  $h^N_{l,f} = [h_{l,f}, \ldots, h_{l-N+1,f}]^T \\\in \mathcal{C}^{N \times 1}$ (see Fig.~\ref{fig:system_model}). To this end, we adopt linear prediction with regressor $v_f \in \mathcal{C}^{N \times 1}$, so that the prediction is given as 
\vspace{-0.3cm}
\begin{align}
\hat{h}_{l+\delta,f} = v_f^\dagger h^N_{l,f}.
\label{eq:predicted_channel}
\\[-0.8cm]\nonumber
\end{align}
The focus on linear prediction is justified by the optimality of linear estimation for Gaussian stationary processes \cite{gallager2008principles}, which provide standard models for fading channels in rich scattering environments \cite{3gpp_tr_901}.

To train the linear regressor, for any frame, we have available the training set $\mathcal{Z}^\text{tr}_f = \{ (x_{i,f}, y_{i,f})\}_{i=1}^{L^\text{tr}} \equiv \{(h_{l,f}^N,  h_{l+\delta,f})\\\}_{l=N}^{L^\text{tr}+N-1}$ encompassing $L^\text{tr}$ input-output examples for the predictor \eqref{eq:general_ridge_sol}. Note that data set $\mathcal{Z}_f^\text{tr}$ can be constructed from $L^\text{tr}+N+\delta-1$ channels $\{ h_{1,f}, \ldots, h_{L^\text{tr}+N+\delta-1,f} \}$. In practice, these channel vectors $h_{l,f}$ are estimated using pilot symbols, and estimation noise can be easily incorporated in the model. We define the corresponding $L^\text{tr} \times N$ input matrix $X_f^\text{tr} = [x_{1,f}^\dagger,\ldots,x_{L^\text{tr},f}^\dagger]^\top$ and $L^\text{tr} \times 1$ target vector $y_f^\text{tr}=[y_{1,f},\ldots,y_{L^\text{tr},f}]^\dagger$, so that the data set can be expressed as the pair $\mathcal{Z}_f^\text{tr} = (X_f^\text{tr}, y_f^\text{tr})$.

Based on the training data $\mathcal{Z}_f^\text{tr}$, for fixed hyperparameters $(\lambda, \bar{v})$, an optimized predictor \eqref{eq:general_ridge_sol} is obtained via the ridge regression problem \eqref{eq:general_ridge} with $X_f^\text{tr}$ and the $y_f^\text{tr}$ in lieu of $X$ and $y$. This yields the regressor $v_\lambda^*(\mathcal{Z}_f^\text{tr}|\bar{v})$. 


\vspace{-0.42cm}
\section{Fast Training via Offline Meta-Learning} 
\label{sec:meta}
\vspace{-0.35cm}
An important problem with the linear predictor described in Sec.~\ref{sec:conven_channel_pred} is that it can only rely on data from the same frame, which may be limited due to the spectral efficiency reduction caused by the transmission of pilot symbols. In this section, we propose a solution to this problem by leveraging data from different frames via offline meta-learning, where such data is available at once in a batch manner. The next section will explore an online solution.
\vspace{-0.55cm}
\subsection{Meta-Training}
\vspace{-0.25cm}
Meta-learning operates in two distinct phases: meta-training and meta-testing. During meta-training, we have access to channel data for $F$ frames, with $L+N+\delta-1$ channels $\{ h_{1,f}, \ldots, h_{L+N+\delta-1,f} \}$ available for each frame, with $L > L^\text{tr}$.
Following the standard meta-learning approach (e.g., \cite{jose2021information_bound}), we split the available $L$ data points into $L^\textrm{tr}$ training pairs $\{(x_{i,f}, y_{i,f})\}_{i=1}^{L^\textrm{tr}} \equiv \{(x_{i,f}^\textrm{tr}, y_{i,f}^\textrm{tr})\}_{i=1}^{L^\textrm{tr}}=\mathcal{Z}_f^\text{tr}$ and $L^\textrm{te}=L-L^\textrm{tr}$ test pairs $ \{(x_{i,f}, y_{i,f})\}_{i=L^\textrm{tr}+1}^{L} \equiv \{(x_{i,f}^\textrm{te}, y_{i,f}^\textrm{te})\}_{i=1}^{L^\textrm{te}}=\mathcal{Z}_f^\text{te}$, resulting in two separate data sets $\mathcal{Z}_f^\text{tr}$ and $\mathcal{Z}_f^\text{te}$. The covariate vectors $\{ x_{i,f} \}$ and labels $\{ y_{i,f} \}$ are defined as described in Sec.~\ref{sec:conven_channel_pred} as a function of the prediction window $N$ and prediction lag $\delta$.
We correspondingly define the $L^\text{tr} \times N$ input matrix $X_f^\text{tr}$ and the $L^\text{tr} \times 1$ target vector $y_f^\text{tr}$, as well as the $L^\text{te} \times N$ input matrix $X_f^\text{te}$ and the $L^\text{te} \times 1$ target vector $y_f^\text{te}$.

We propose to optimize the hyperparameter vector $\bar{v}$, for a fixed $\lambda > 0$, based on meta-training data. This is done so as to minimize the sum-loss of the predictors defined by the trained regressors $v_\lambda^{*}(\mathcal{Z}^\text{tr}_f|\bar{v})$ across all frames $f=1,\ldots,F$. Accordingly, estimating the loss in each frame $f$ via the test set $\mathcal{Z}^\text{te}_f$ yields the meta-training problem
\vspace{-0.3cm}
\begin{align}
\bar{v}^*_\lambda &= \argmin_{\bar{v}} \left\{ \sum_{f=1}^F \sum_{i=1}^{L^\textrm{te}}\big|v_{\lambda}^*(\mathcal{Z}_f^\textrm{tr}|\bar{v})^\dagger x_{i,f}^\textrm{te} - y_{i,f}^\textrm{te}\big|^2 \right\}.
\label{eq:meta_objective}
\\[-0.8cm]\nonumber
\end{align}

As studied in \cite{denevi2018learning}, the minimization in \eqref{eq:meta_objective} is a least squares problem that can be solved in closed form as
\vspace{-0.3cm}
\begin{align}
    \bar{v}^*_\lambda &= \argmin_{\bar{v}} \sum_{f=1}^F \norm{\tilde{X}_f^\text{te}\bar{v} - \tilde{y}_f^\text{te}}^2 = (\tilde{X}^\dagger \tilde{X})^{-1} \tilde{X}^\dagger \tilde{y},\label{eq:closed_from_sol_meta_scalar}
\\[-0.9cm]\nonumber
\end{align}
where $L^\text{te} \times N$ matrix $\tilde{X}_f^\text{te}$ contains by row the Hermitian transpose of the $N \times 1$ pre-conditioned input vectors $\{ \lambda (A_f^\text{tr})^{-1} x_{i,f}^\text{te} \}_{i=1}^{L^\text{te}}$, with $A_f^\text{tr} = (X_f^\text{tr})^\dagger X_f^\text{tr} + \lambda I$,; $\tilde{y}_f^\text{te}$ is $L^\text{te} \times 1$ vector containing vertically the complex conjugate of the transformed outputs 
$\{ (y^\text{te}_{i,f} - (y_f^\text{tr})^\dagger X_f^\text{tr} (A_{f}^\text{tr})^{-1} x_{i,f}^\text{te} \}_{i=1}^{L^\text{te}}$;
the $F L^\text{te} \times N$ matrix $\tilde{X} = [\tilde{X}_1^\text{te}, \ldots,\tilde{X}_F^\text{te}]^\top$ stacks vertically the $L^\text{te} \times N$ matrices $\{\tilde{X}_f^\text{te}\}_{f=1}^F$; and the $F L^\text{te} \times 1$ vector $\tilde{y} = [\tilde{y}_{1}^\text{te},\ldots,\tilde{y}_{F}^\text{te}]^\top$ stacks vertically the $L^\text{te}\times 1$ vectors $\{\tilde{y}_f^\text{te}\}_{f=1}^F$.

Unlike standard meta-learning algorithms used by most papers on communications \cite{park2020learning,cohen2021learning,raviv2021meta, park2020meta, jiang2019mind,yuan2020transfer}, such as model-agnostic meta-learning (MAML), the proposed meta-learning procedure is not based on iterative processing such as gradient descent, and it does not require the tuning of hyperparameters such as a learning rate schedule. We will discuss below how to set the only free hyperparameter $\lambda$. 

Reference \cite{denevi2018learning} utilizes an additional validation dataset to optimize the scalar hyperparameter $\lambda$ via a cross-validation approach \cite{simeone2018brief}. Given that this would require the transmission of additional frames or prevent the full use of the available frames for meta-learning, we propose to tune $\lambda$ by addressing the problem
\vspace{-0.37cm}
\begin{align}
    \lambda^* = \argmin_{\lambda \in \Lambda} \left\{ \sum_{f=1}^F \sum_{i=1}^{L^\textrm{te}}\big|v_{\lambda}^*({\mathcal{Z}}_f^\textrm{tr}|\bar{v}^*_\lambda)^\dagger {x}_{i,f}^\textrm{te} - {y}_{i,f}^\textrm{te}\big|^2 \right\},\label{eq:tuning_lambda}
    \\[-0.9cm]\nonumber
\end{align}
over a predefined discrete search $\Lambda$ space for $\lambda$. In order to mitigate overfitting that may result from optimizing both hyperparameters $\bar{v}$ and $\lambda$ based on the same split $(\mathcal{Z}_f^\text{tr},\mathcal{Z}_f^\text{te})$ of data set $\mathcal{Z}_f$, we have implemented \eqref{eq:tuning_lambda} by choosing a different random split of the data set.


\vspace{-0.45cm}
\subsection{Meta-Testing}
\vspace{-0.25cm}
After meta-learning, during runtime, for a new frame $f^{\textrm{new}}$, we observe training channels $h_{l,f^\textrm{new}}$ with $l=1,\ldots,L^{\text{new}}+N+\delta-1$. Accordingly, we have the training set $\mathcal{Z}_{f^\text{new}} = \{(h_{l,f^\text{new}}^N, h_{l+\delta,f^\text{new}})\}_{l=N}^{L^\text{new}+N-1}$. Based on these data and on the meta-trained hyperparameters $(\lambda^*,\bar{v}_{\lambda^*}^*)$, we train a channel predictor via ridge regression \eqref{eq:general_ridge}, obtaining 
\vspace{-0.2cm}
\begin{align}
v_{f^\textrm{new}}^* = v_{\lambda^*}^*(\mathcal{Z}_{f^\text{new}}|\bar{v}_{\lambda^*}^*).
\label{eq:mte_scalar}
\\[-0.6cm]\nonumber
\end{align}

\vspace{-0.75cm}
\section{Fast Training via Online Meta-Learning} 
\vspace{-0.35cm}
\label{sec:online_meta}
In this section, we briefly consider a scenario in which frames are transmitted in a streaming fashion so that, at any current frame $f$, we can use as meta-training data the channel data from past frames. Specifically, we assume that at frame $f$, we have access to data $\{\{ h_{l,f'} \}_{l=1}^{L+N+\delta-1}\}_{f'=\min{(f-M,1)}}^{f-1}$ from the most recent $M$ frames.

In order to facilitate an online update of the hyperparameter vector $\bar{v}$ as more frames are collected, we develop an iterative gradient-based solution that follows EP \cite{scellier2017equilibrium}. 

To start, we define $L_{f'}(v) := ||X_{f'}^\text{tr}v-y_{f'}^\text{tr}||^2$ for the training loss in frame $f'$; $L_{f'}^\text{inner}(v|\bar{v}):= L_{f'}(v)+\lambda\norm{v-\bar{v}}^2$ for the corresponding regularized loss; and $L_{f'}^\text{outer}(v) := ||X_{f'}^\text{te}v-y_{f'}^\text{te}||^2$ for the corresponding test loss. With these definitions, the optimal hyperparameter vector \eqref{eq:meta_objective} obtained at frame $f$ based on the observation of data from frames $f'=\max{(f-M,1)},\ldots,f-1$ can be rewritten as
\vspace{-0.35cm}
\begin{align}
    \nonumber
    \bar{v}_{\lambda,M}^*(f) = \argmin_{\bar{v}} &\sum_{f'=\max{(f-M,1)}}^{f-1}  L_{f'}^\text{outer}\left(v_\lambda^*(\mathcal{Z}_{f'}^\text{tr}|\bar{v})\right), \\ \text{ where } &v_\lambda^*(\mathcal{Z}_{f'}^\text{tr}|\bar{v}) = \argmin_{v} L_{f'}^\text{inner}(v|\bar{v}).
    \label{eq:obj_for_gradient_methods} \\[-30pt]\nonumber
\end{align}
EP tackles problem \eqref{eq:obj_for_gradient_methods} via gradient descent \cite{simeone2018brief} by introducing total loss function
\vspace{-0.2cm}
\begin{align}
    L_{f'}^\text{total}(v|\bar{v},\alpha) = L_{f'}^\text{inner}(v|\bar{v}) + \alpha L_{f'}^\text{outer}(v),
\end{align}
with an additional real-valued hyperparameter $\alpha \in \mathbb{R}$. Defining
\vspace{-0.2cm}
\begin{align}
 \label{eq:stationary_points}
v&_{f'}^{\alpha,\bar{v}} = \argmin_v L_{f'}^\text{total}(v|\bar{v},\alpha)\\\nonumber
&=\argmin_{v} \left\{ \norm{X_{f'}^\text{tr}v - y_{f'}^\text{tr}}^2 + \alpha \norm{X_{f'}^\text{te}v - y_{f'}^\text{te}}^2 + \lambda \norm{v-\bar{v}}^2\right\} \\
    &= (A_{f'}^\alpha)^{-1}\left((X_{f'}^\alpha)^\dagger y_{f'}^\alpha + \lambda \bar{v}\right), \text{ with } A_{f'}^\alpha = \left((X_{f'}^\alpha)\right)^\dagger X_{f'}^\alpha + \lambda I,
    \nonumber\\[-0.7cm]\nonumber
\end{align}
where the $L \times N$ matrix $X_{f'}^\alpha = [X_{f'}^\text{tr}, \sqrt{\alpha} X_{f'}^\text{te}]^\top$ stacks vertically the training input matrix $X_{f'}^\text{tr}$ and the weighted test input matrix $\sqrt{\alpha} X_{f'}^\text{te}$; and the $L\times 1$ vector $y_{f'}^\alpha = [y_{f'}^\text{tr},\sqrt{\alpha} y_{f'}^\text{te}]^\top$ stacks vertically the training target vector $y_{f'}^\text{tr}$ and the weighted test target vector $\sqrt{\alpha} y_{f'}^\text{te}$. Based on definition \eqref{eq:stationary_points}, we have $v_{f'}^{\alpha=0,\bar{v}} = v_\lambda^*(\mathcal{Z}^\text{tr}_{f'}|\bar{v})$, which is the solution of the inner problem in \eqref{eq:obj_for_gradient_methods}. Furthermore, it can be proved \cite{scellier2017equilibrium} that the gradient $\nabla_{\bar{v}}L_{f'}^\text{outer}(v_\lambda^*(\mathcal{Z}_{f'}^\text{tr}|\bar{v}))$ of the objective of the meta-training problem \eqref{eq:obj_for_gradient_methods} can be expressed as the limit
\vspace{-0.1cm}
\begin{align}
\nonumber
\nabla_{\bar{v}}&L_{f'}^\text{outer}\left(v_\lambda^*(\mathcal{Z}_{f'}^\text{tr}|\bar{v})\right)\\\nonumber&= \lim_{\alpha \rightarrow 0} \frac{1}{\alpha} \left( \frac{\partial L_{f'}^\text{total}}{\partial \bar{v}}\left(v_{f'}^{\alpha,\bar{v}}|\bar{v},\alpha\right) - \frac{\partial L_{f'}^\text{total}}{\partial \bar{v}}\left(v_{f'}^{0,\bar{v}}|\bar{v},0\right)  \right) \\\label{eq:EP_essential}
&=\lim_{\alpha \rightarrow 0} \frac{1}{\alpha} \left(2\lambda(v_{f'}^{0,\bar{v}}-v_{f'}^{\alpha,\bar{v}})\right).\\[-0.9cm]\nonumber
\end{align}
With a small enough scalar $\alpha \neq 0$, regardless of its sign, we can hence estimate the gradient $\nabla_{\bar{v}}L_{f'}^\text{outer}\left(v_\lambda^*(\mathcal{Z}_{f'}^\text{tr}|\bar{v})\right)$ as
\vspace{-0.3cm}
\begin{align}
    \nonumber
    \frac{1}{\alpha} \big(2\lambda(v_{f'}^{0,\bar{v}}-v_{f'}^{\alpha,\bar{v}})&\big) = 
    \frac{1}{\alpha} \bigg(2\lambda\Big((A_{f'}^\text{tr})^{-1}\left((X_{f'}^\text{tr})^\dagger y_{f'}^\text{tr} + \lambda \bar{v}\right)\\&-(A_{f'}^\alpha)^{-1}\left((X_{f'}^\alpha)^\dagger y_{f'}^\alpha + \lambda \bar{v}\right)\Big)\bigg),
    \label{eq:EP_grad_actual}
    \\[-0.9cm]\nonumber
\end{align}
where we have used \eqref{eq:stationary_points}. In order to obtain an online learning rule, we propose to apply gradient descent using gradient estimate \eqref{eq:EP_grad_actual} and adopting as initialization the final iterate produced at frame $f-1$.

\vspace{-0.4cm}
\section{Experiments}
\label{sec:exp}
\label{subsec:numerical_results}
\vspace{-0.3cm}
In this section, we present a number of experiments for both offline and online meta-learning\footnote{Code for regenerating the results of this paper can be found at \url{https:github.com/kclip/scalar-channel-meta-prediction}}.

\noindent\emph{Random Doppler Frequency (Offline)}:
\label{subsubsec:toy_numerical}
\begin{figure}
    \centering
    \hspace{-0.7cm}
    \includegraphics[width=0.85\columnwidth]{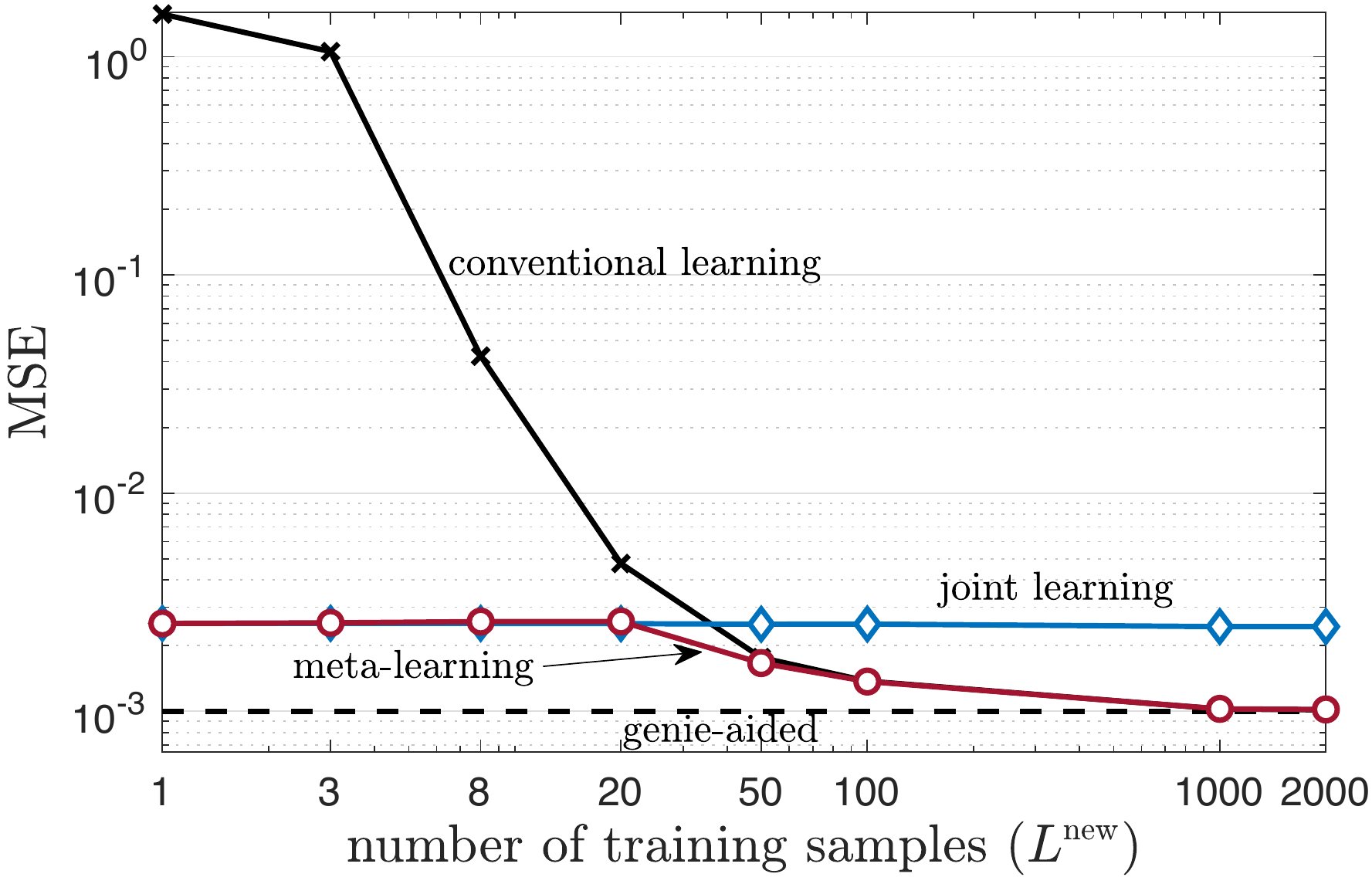}
    \vspace{-0.26cm}\caption{Channel prediction performance as a function of the number of training samples $L^\text{new}$, with window size $N=5$ and lag $\delta=3$ for the setting with random maximum Doppler frequencies.}
    \label{fig:scalar_pred_not_toy}
    \vspace{-0.55cm}
\end{figure}
Consider the Doppler spectra following Jakes' model 
\vspace{-0.3cm}
\begin{align}
S_1(w)=\frac{1}{\frac{w_m}{2} \sqrt{1-(w/w_m)^2}}, \textrm{ } |w| \leq w_m
\end{align}
and the rounded model \cite{hari2003channel}
\vspace{-0.3cm}
\begin{align}
\nonumber
S_2(w) &= \alpha \Big[ a_0 + a_2(w/w_m)^2 + a_4(w/w_m)^4 \Big], \textrm{ } |w| \leq w_m, \\ &\quad\quad\quad\quad\textrm{ with } \alpha = \frac{1}{\frac{w_m}{\pi}\big[ a_0 + a_2/3 + a_4/5 \big]}.
\end{align}
We consider $a_0=1, a_2=-1.72, a_4=0.785$ \cite{hari2003channel}. The Doppler spectrum $S_f(w)$ in each frame $f$ is assumed to be generated by first picking at random, with the same probability, a Jakes' or rounded spectrum; and then selecting a random Doppler frequency ${w_m}/{2\pi} \sim \mathcal{U}_{[0.05, 0.1]}$. We fix $F=80$ meta-training frames with number of slots per frame $L = L^\text{tr} + L^\text{te} = L^\text{new} + 100$.

We compare the proposed meta-learning scheme to \emph{(i)} \emph{conventional learning}, which directly applies the regressor \eqref{eq:mte_scalar} with $\lambda = 0$; \emph{(ii)} \emph{genie-aided prediction}, which implements LMMSE strategy based on knowledge of the Doppler spectrum;  \emph{(iii}) \emph{joint learning}, which utilizes previous channel data at frames $f$ jointly with the current channel at frame $f^\textrm{new}$ to obtain the regressor $v^*_{f^\textrm{new}}$ by solving the problem 
\vspace{-0.2cm}
\begin{align}
\min\limits_{{v}} \Big(
\sum\limits_{f}\sum\limits_l &|| v^\dagger h_{l,f}^N - h_{l+\delta,f}  ||^2 \nonumber\\[-8pt]&+ \sum\limits_l || v^\dagger h_{l,f^\textrm{new}}^N -h_{l+\delta,f^\textrm{new}} ||^2\Big). \\[-25pt]\nonumber
\end{align}
Conventional learning and joint learning schemes correspond to LMMSE strategies based on covariance matrices estimated using pilots received, respectively, only in the new frame or in all the previous frames along with the new frame.

In Fig.~\ref{fig:scalar_pred_not_toy}, we compute the mean squared error (MSE) averaged over $10,000$ samples. For prediction, we consider window size $N=5$ with lag size $\delta=3$. 
During meta-testing, we compute the channel prediction MSE with varying number of training  samples $L^\text{new}$ for the new frame $f^\textrm{new}$ as shown in Fig.~\ref{fig:scalar_pred_not_toy}. All the considered schemes except for joint learning are seen to converge to the optimal performance as $L^\text{new}$ grows, while meta-learning is observed to predict future channels even only with a single sample, i.e., with $L^\text{new}=1$.
Additionally, we also verified (not shown) that deep neural network-based predictors such as recurrent neural network or multilayer perceptron could not outperform the linear predictors considered in this work, as also reported in \cite{kim2020massive, jiang2020long}.

\begin{figure}
    \centering
    \hspace{-0.4cm}
    \includegraphics[width=0.8\columnwidth]{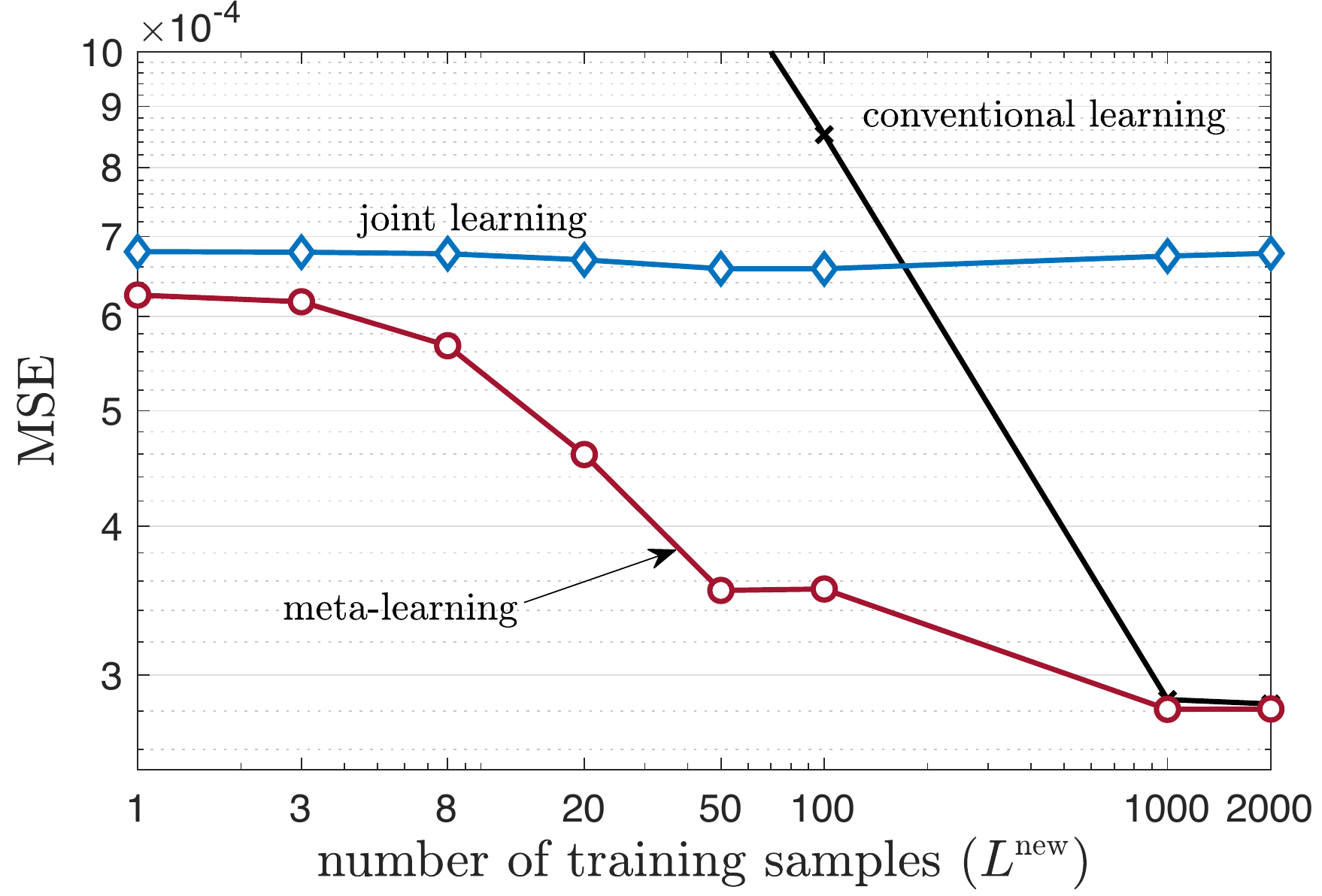}
    \vspace{-0.26cm}\caption{Channel prediction performance as a function of the number of training samples $L^\text{new}$, with window size $N=5$ and lag $\delta=3$ under single clustered CDL-C model with random user velocity at each frame $f$ \cite{3gpp_tr_901}.}
    \label{fig:scm_over_supp}
    \vspace{-0.3cm}
\end{figure}

\noindent\emph{Standard Channel Model (Offline)}: \label{sec:exp_standard_off}
Here, we evaluate the proposed solution under the 5G standard spatical channel model (SCM) model \cite{3gpp_tr_901}. To obtain frequency-flat fading, from the CDL-C channel model \cite{3gpp_tr_901}, we take only the first cluster into account to generate the channel model $h_{l,f}$. The Doppler frequency for each path is randomly scaled to have values between ${w_{d,f}}/{2\pi} \in [10,100]$ by randomly changing the velocity of the user. Fig.~\ref{fig:scm_over_supp} shows channel prediction performance with the same setting as Fig.~\ref{fig:scalar_pred_not_toy}, while considering $20$ dB of channel estimation noise with $P=100$ pilots. We observe that the proposed meta-learning approach works well even for channel models that do not strictly follow stationary Gaussian processes.

\begin{figure}
    \centering
    \hspace{-0.7cm}
    \includegraphics[width=0.85\columnwidth]{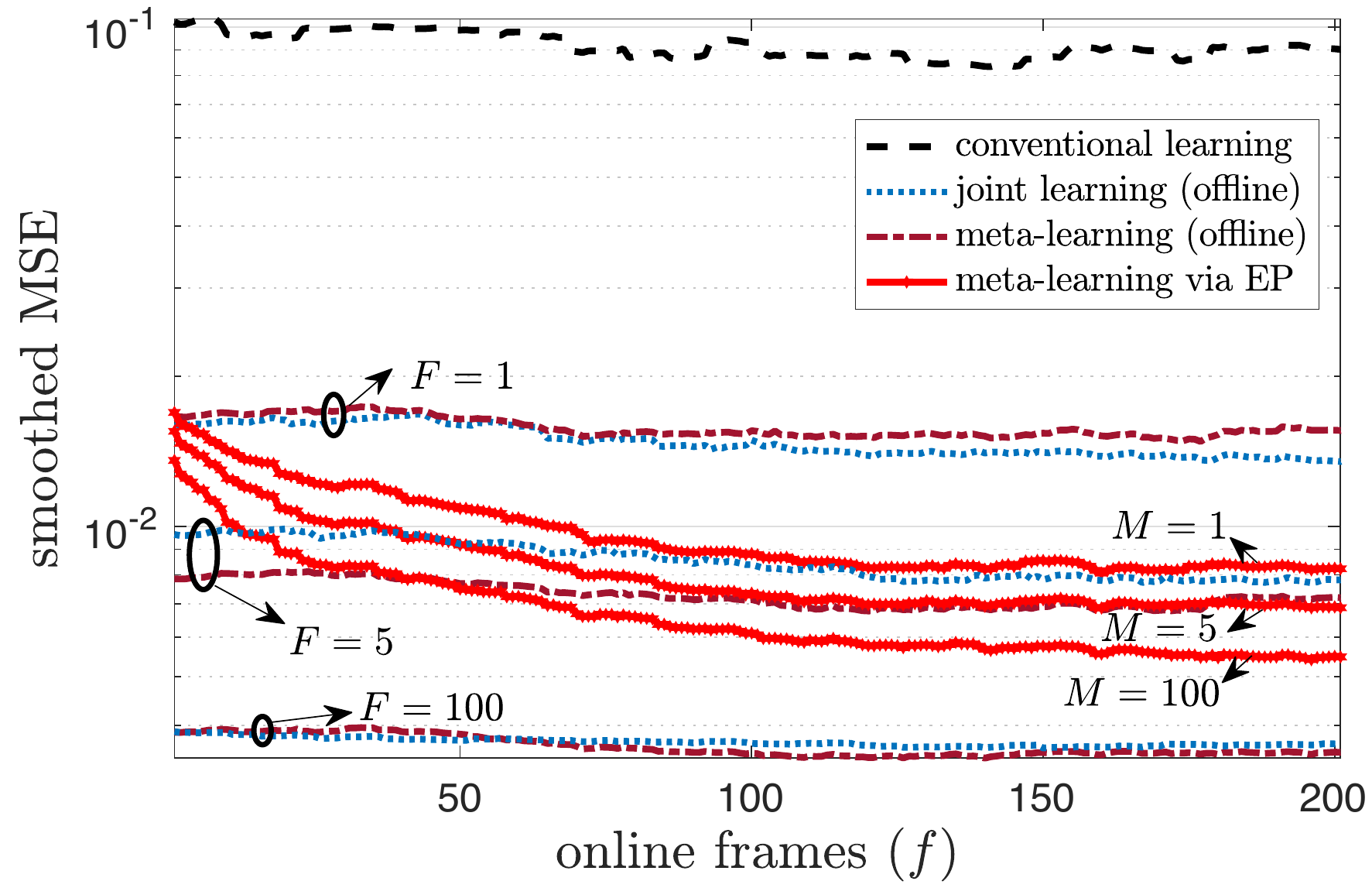}
    \vspace{-0.4cm}\caption{Channel prediction performance as a function of current frame index $f$ with window size $N=5$ and lag $\delta=3$ under single clustered CDL-C model with random user velocity for each frame $f$ \cite{3gpp_tr_901}. MSE is averaged over $10$ independent experiments and smoothed over $100$ frames.}
    \label{fig:online_meta}
    \vspace{-0.5cm}
\end{figure}

\noindent\emph{Gradient-Based Meta-Learning  (Online)}:
\label{sec:exp_standard_on}
Lastly, we consider an online scenario in which frames are observed in a streaming fashion. We assume same standard channel model as in Sec.~\ref{sec:exp_standard_off} during frames $f=1,\ldots,300$ and compute the MSE at each frame $f$ with fixed lambda $\lambda =1$. We average over $10$ independent experiments and plot the MSE smoothed with moving average with length $100$. Fig.~\ref{fig:online_meta} shows that online meta-learning via EP outperforms conventional learning and offline joint learning even with $M=1$, with the MSE decreasing as $M$ is increased.

\vspace{-0.5cm}
\section{Conclusions}
\vspace{-0.36cm}
This paper considered linear channel prediction via meta-learning for both offline and online scenarios. The proposed meta-learned linear filters via closed-form expression and EP, were seen to effectively reduce the pilot overhead needed in each frame. Future work may consider prediction for frequency-selective MIMO fading channel.

\vfill\pagebreak

\bibliographystyle{IEEEbib}
\bibliography{ref}

\end{document}